\documentclass[preprint]{aastex61}

\usepackage{amsmath,amssymb,amstext}

\usepackage{epstopdf}

\usepackage{amsmath,amssymb,amstext}

\usepackage{natbib}
\usepackage{color}
\definecolor{darkred}{RGB}{175,0,0}

\makeatletter

\newcommand{\Rmnum}[1]{\expandafter\@slowromancap\romannumeral #1@}
\makeatother

\begin{document}

\title{Feature ranking of active region source properties in solar flare forecasting and the uncompromised stochasticity of flare occurrence}

\author{Cristina Campi}
\affil{Dipartimento di Matematica "Tullio Levi Civita", Universit\`a di Padova, via Trieste 63 35121 Padova, Italy}
\email{cristina.campi@unipd.it}

\author{Federico Benvenuto}
\affil{Dipartimento di Matematica, Universit\`a di Genova, via Dodecaneso 35 16146 Genova, Italy}
\email{benvenuto@dima.unige.it}

\author{Anna Maria Massone}
\affil{Dipartimento di Matematica, Universit\`a di Genova, via Dodecaneso 35 16146 Genova, Italy}
\affil{CNR - SPIN Genova, via Dodecaneso 33 16146 Genova, Italy}
\email{massone@dima.unige.it}

\author{D. Shaun Bloomfield} 
\affil{Department of Mathematics, Physics and Electrical Engineering, Northumbria University, Newcastle upon Tyne, NE1 8ST, UK}
\email{shaun.bloomfield@northumbria.ac.uk}

\author{Manolis K. Georgoulis}
\affil{Department of Physics and Astronomy, Georgia State University, Atlanta (GA), USA}
\affil{RCAAM of the Academy of Athens, Athens, Greece}
\email{manolis.georgoulis@academyofathens.gr}

\author{Michele Piana}
\affil{Dipartimento di Matematica, Universit\`a di Genova, via Dodecaneso 35 16146 Genova, Italy}
\affil{CNR - SPIN Genova, via Dodecaneso 33 16146 Genova, Italy}
\email{piana@dima.unige.it}

%\begin{abstract}
%
%\end{abstract}

%\maketitle

\begin{abstract}
Solar flares originate from magnetically active regions but not all solar active regions give rise to a flare. Therefore, the challenge of solar flare prediction benefits by an intelligent computational analysis of physics-based properties extracted from active region observables, most commonly line-of-sight or vector magnetograms of the active-region photosphere. For the purpose of flare forecasting, this study utilizes an unprecedented 171 flare-predictive active region properties, mainly inferred by the {\em{Helioseismic and Magnetic Imager}}
onboard the {\em{Solar Dynamics Observatory (SDO/HMI)}} in the course of the European Union Horizon 2020 FLARECAST project. Using two different supervised machine learning methods that allow feature ranking as a function of predictive capability, we show that: i) an objective training and testing process is paramount for the performance of every supervised machine learning method; ii) most properties include overlapping information and are therefore highly redundant for flare prediction; iii) solar flare prediction is still - and will likely remain - a predominantly probabilistic challenge.
\end{abstract}


\section{Introduction}
Solar flares are the most explosive events in the heliosphere, releasing in an abrupt way up to $10^{33}$ ergs of energy in a time interval typically ranging between $10$ and $1000$ seconds \citep{be17}. This energy is previously stored in specific magnetic configurations and, when magnetic reconnection occurs \citep{prfo02}, it is transformed into mass acceleration, heating and electromagnetic radiation at all wavelengths. It is also established that flares are a major space weather agent in the heliosphere \citep{sc06}, while, as secondary effects through their correlation with coronal mass ejections, they induce geospace and ionospheric disturbances, malfunctions and impairments on technologies in the geosphere, such as flight navigation, satellite communication and power grid distribution. 

Solar flare forecasting is a prominent discipline \citep{gaetal02,geru07,sc07,lietal07,bale08,waetal08,lietal08,yuetal09,coqa09,ahetal13,boco15,baetal16,sako17,muetal17,mietal17,beetal18,huetal18,puetal18,mapi18,nietal18} within the recent field of space weather forecasting that relies on the availability of two ingredients; one observational and one computational. First, it is well-established that solar active regions (ARs) exclusively host major flares and therefore flare prediction needs experimental data on AR properties, {\bf{associated to the photospheric and coronal magnetic field; however, coronal information has only recently started being used in the form of EUV images given as input to a deep learning network by \citet{nietal18}.}} Second, this information on AR magnetic properties can be processed for prediction purposes by means of a computational method for data analysis; machine learning has recently offered strong candidates for such methods. 

Since February 2010, the {\em{Helioseismic and Magnetic Imager}} onboard the {\em{Solar Dynamics Observatory}} ({\em{SDO/HMI}}) \citep{scetal12} is providing both line-of-sight and vector magnetograms of the full solar disk at a (vector magnetogram) cadence of $12$ minutes. {\em{SDO/HMI}} magnetograms can be used for solar flare prediction according to two different approaches. {\bf{First, HMI magnetograms are utilized to calculate a variety of properties either from the line-of-sight component only, from the radial component only, or from all three vector components. Various single-valued quantities, hereafter referred to as features, can be calculated from these property images through a variety of techniques (e.g., thresholding, feature recognition, etc), such that calculation of one physical property may provide multiple features as inputs to machine learning (i.e., image maximum, total, and moments). Of course, additional features that are not derived from property images may also contribute to the input dataset}}. Second, from a deep learning perspective, {\em{HMI}} images can be given as input to Convolutional Neural Networks (CNNs) that automatically perform a probabilistic forecasting. This present paper follows the first approach, and this is for several reasons. First, we had at disposal the {\bf{property}} extraction power provided by the algorithms developed within the Horizon 2020 FLARECAST project (http://flarecast.eu), which generated datasets of almost $200$ features {\bf{determined from properties extracted from photospheric {\em{SDO/HMI}} vector magnetograms}}. This database of features probably represents the highest data dimensionality currently available for flare forecasting purposes. Second, one of the objectives of our research was to determine to what extent AR properties are redundant when forecasting flares, and a straightforward way to do this is by ranking the extracted features according to their predictive capability. Finally, so far most publications in flare prediction utilize feature-based machine learning methods, so another objective of this paper is to investigate how data preparation (and, specifically, the preparation of the training set in the case of supervised algorithms) impacts the prediction scores. Specifically, the analysis performed in this paper relies on two supervised machine learning algorithms that combine prediction with feature ranking, namely hybrid LASSO \citep{beetal18} and Random Forest \citep{br01}. However, two other methods of this kind, namely logit \citep{wuetal09} and a support vector machine for classification \citep{cova95}, have been also applied to the same datasets for verification purposes, with coherent results.

The content of the paper is as follows. Section 2 overviews the data analysis procedure, describing in detail the features used for prediction, the data preparation process, and key aspects of the machine learning methods adopted. Section 3 contains the results of the analysis, while Section 4 discusses these results. Our conclusions are offered in Section 5.

\section{Methods}

\subsection{Data and features}
Our analysis relies on the Near-Realtime (NRT) Space Weather HMI Archive Patch (SHARP) data product of the HMI database \citep{boetal14}. These data comprise 2D images of continuum intensity, the full three-component magnetic field vector, and the line-of-sight component of each photospheric HARP. We then made use of property extraction algorithms developed by the FLARECAST Consortium in order to construct a property database made of property vectors comprising up to $171$ components. FLARECAST algorithms first extracted the following $167$ features, often duplicating the property calculation step on $B_{los}$ and $B_{radial}$ input data, as per the findings of \citet{guetal18}: 
\begin{itemize}
\item Schrijver's R value \citep{sc07}: $1$ property yielding a total of $2$ features.
\item Multifractal structure and function spectrum on a 2D image: $2$ properties yielding a total of $4$ features.
\item Falconer's total free magnetic energy proxy $WL_{SG}$ \citep{faetal08}: $1$ property yielding a total of $2$ features.
\item Sum of the horizontal magnetic gradient, $G_S$, and the separation of opposite-polarity sunspot subgroups, $S_{l-f}$ \citep{koer16}: $2$ properties yielding a total of $4$ features.
\item Spectral power indices extracted by means of the Fourier transform and of a continuous wavelet transform: $1$ property yielding a total of $4$ features.
\item Magnetic polarity inversion line (MPIL) characteristics: $3$ properties yielding a total of $6$ features.
\item Effective connected magnetic field strength ($B_{eff}$ ): $1$ property yielding a total of $2$ features.
\item Vertical decay index of potential field: $4$ properties yielding a total of $8$ features.
\item Non-neutralized electric currents: $1$ property yielding $1$ feature.
\item Ising energy ($E$): $1$ property yielding a total of $4$ features.
\item Fractal dimension ($D$): $1$ property yielding a total of $2$ features.
\item Flow field characteristics: $6$ properties yielding a total of $16$ features.
\item Magnetic helicity and energy injection rate: $14$ properties yielding a total of $14$ features.
\item SHARP keywords calculated from their corresponding vector and line-of-sight magnetograms: $16$ properties yielding a total of $100$ features (including the maximum, total, median, mean, standard deviation, skewness and kurtosis over the SHARP field-of-view).
\end{itemize}
{\bf{SHARP ARs are associated to solar flares of GOES class C1 and above (C1+) and solar flares of GOES class M1 and above (M1+) by means of a standard procedure. It is first verified whether the SHARP data contain NOAA-numbered regions (i.e., sunspot groups) by comparison with NOAA's daily Solar Region Summary (SRS) file immediately before the SHARP observations. Then, if any NOAA number is assigned to the SHARP data, the process searches the NOAA/SWPC daily events lists for GOES flares occurring in the same source region during the entire disk passage. Once the flare association is realized, the following four details become available for all flares and these are used in assigning flare outcome labels:}}
\begin{itemize}
\item GOES {\bf{peak}} magnitude ($F_M$).
\item Time difference (in seconds) between the SHARP observation time and the flare start time ($\tau_s$).
\item Time difference (in seconds) between the SHARP observation time and the flare peak time ($\tau_p$).
\item Time difference (in seconds) between the SHARP observation time and the flare end time ($\tau_e$).
\end{itemize}
Eventually, this analysis provides $167$ {\bf{features}} extracted from the HMI images. Four further features come from the NOAA/SRS database: the mean heliographic longitude and latitude of each AR, a binary label encoding the presence of a flare in the past $24$ hours and the flare index of events occurring within the past $24$ hours. {\bf{A summary of all resulting features considered in the analysis can be found at both the url https://api.flarecast.eu/property/ui/ and in Tables 1-3 in the Appendix.}}

\subsection{Data preparation}
The experiment designed in this paper relies on supervised machine learning, which requires appropriate historical sets to train the prediction networks. To enforce consistency in time and robustness of our tests, we constructed four training sets, each one corresponding to a specific forecast issuing time expressed as Universal Time [UT], namely 00:00, 06:00, 12:00, and 18:00. For each issuing time we considered the set of {\em{SDO/HMI}} images recorded at that time in the range of days between 14 September 2012 and 30 April 2016, with $24$-hour sampling. While filling up the training set we took care to focus on ARs rather than on feature vectors. In fact, around $2/3$ ARs were randomly extracted from the set of all ARs belonging to a specific issuing time and the $171$-dimension feature vectors associated to each AR were labelled by annotating whether a GOES C1+ flare occurred in the next $24$ hours. The set of feature vectors associated to the remaining $1/3$ ARs was not labeled and was provided as test set for experiments to supervised learning algorithms trained on the training set. In this manner, training and testing do not overlap in any way, either in time or in terms of ARs examined. We finally point out that, for each issuing time, the random, complete separation of ARs into training and test sets was replicated $100$ times to enable statistical robustness of the results. A similar procedure was implemented to generate training sets to use for the prediction of GOES M1+ flares. {\bf{The reason why we did not consider the prediction of flares with class above X is because they are extremely seldom in the database (less than $0.2\%$ of the overall point-in-time events in the original training set). C1+ and M1+ flares are around $26 \%$ and $4\%$ of the set content, respectively.}}

\subsection{Prediction and feature ranking methods.}
Two different machine learning methods, namely hybrid LASSO (HLA) and Random Forest (RF) are utilized in this paper, for both performing the binary prediction of the flare occurrence and for additionally identifying the effectiveness with which the different features contribute to the prediction. 

LASSO methods \citep{ti96} are intrinsically regression methods and therefore they are not originally conceived for applications that require a binary YES/NO output. However, in \citet{beetal18} a threshold optimization is introduced to the LASSO outcome in order to realize classification by means of fuzzy clustering \citep{beetal84}. The idea of HLA is therefore to use LASSO in the first step in order to promote sparsity and to realize feature selection; this step provides an optimal estimate of the model parameters and corresponding predicted output. In the second step, Fuzzy C-Means is applied for clustering the predicted output in two classes. The main advantage of this approach is in the use of fuzzy clustering to automatically classify the regression output in two classes. Indeed, fuzzy clustering identifies flaring/non-flaring events with a thresholding procedure that is data-adaptive and completely operator-independent. Details about HLA as implemented in the present paper can be found in \citet{beetal18}.

RF belongs to the family of the ensemble methods, i.e. methods that make use of a combination of different learning models to increase the classification accuracy. In particular, RF works as a large collection of de-correlated decision trees. In fact, {\bf{here the training set is randomly divided into $10$ subsets and for each subset a separate decision tree is built}}. Each decision tree is then used to classify an incoming unlabeled sample. If correctly implemented, RF can be used as feature rankers. In fact, the relative depth of a feature used as a decision node in a tree can be identified as the relative importance of that feature with respect to the predictability of the target variable. Features used at the top of the tree contribute to the final prediction decision of a larger fraction of the input samples. The expected fraction of the samples they contribute to can thus be used as an estimate of the relative importance of the features. Details about RF as implemented in the present paper can be found in \citet{br01}.

Once the two machine learning methods have been applied to the input data, predictors are ranked by using Recursive Feature Elimination (RFE). This 
iterative procedure can be summarized as follows:
\begin{enumerate}
\item Train	 the	classifier.
\item  Compute the ranking for all features.
\item Remove the feature with the smallest ranking.
\end{enumerate}
Details about RFE as implemented in the present paper can be found in \citet{guetal02}.
%More	specifically	the	RFE	algorithm	works	in	this	way:
%\begin{mdframed}
%Let $X$ be the	matrix	with	dimension	$N\times P$ whose columns	contain	 the	 feature	 values	 for	each	
%sample	 in	 the	 original	 training set, $y$ the	 $N\times 1$ vector	 of	 the	 corresponding	 labels,  
%$\mathcal{P} = [1,\cdots,P]$ the $P$-dimensional list of all the predictors, and $\mathcal{R} = []$ the list of ranked predictors, initially empty.\\ \\
%While $\mathcal{P}$ is not empty:
%\begin{itemize}
%\item $Xp = X[:,\mathcal{P}]$;
%\item training phase, estimating $\beta = K(Xp,y)$ where $K$ represents the method
%\item rank	the	features	according	to	the ranking criteria $c = [ w_1^2, w_2^2, \cdots, w_P^2];$
%\item find	the	index	of	the	feature	with	smallest	ranking	 $j^* = argmin(c);$
%\item add	the	selected	feature	to	the	set  $\mathcal{R}$: $\mathcal{R} = \mathcal{R} \cup p_{j^*};$ 
%\item remove $p_{j^*}$ from the set  $\mathcal{P}$:  $\mathcal{P}=  \mathcal{P} \backslash p_{j^*};$
%\end{itemize}
%\end{mdframed}

\section{Results}

\begin{figure}
\begin{center}
 \includegraphics[width=10.5cm]{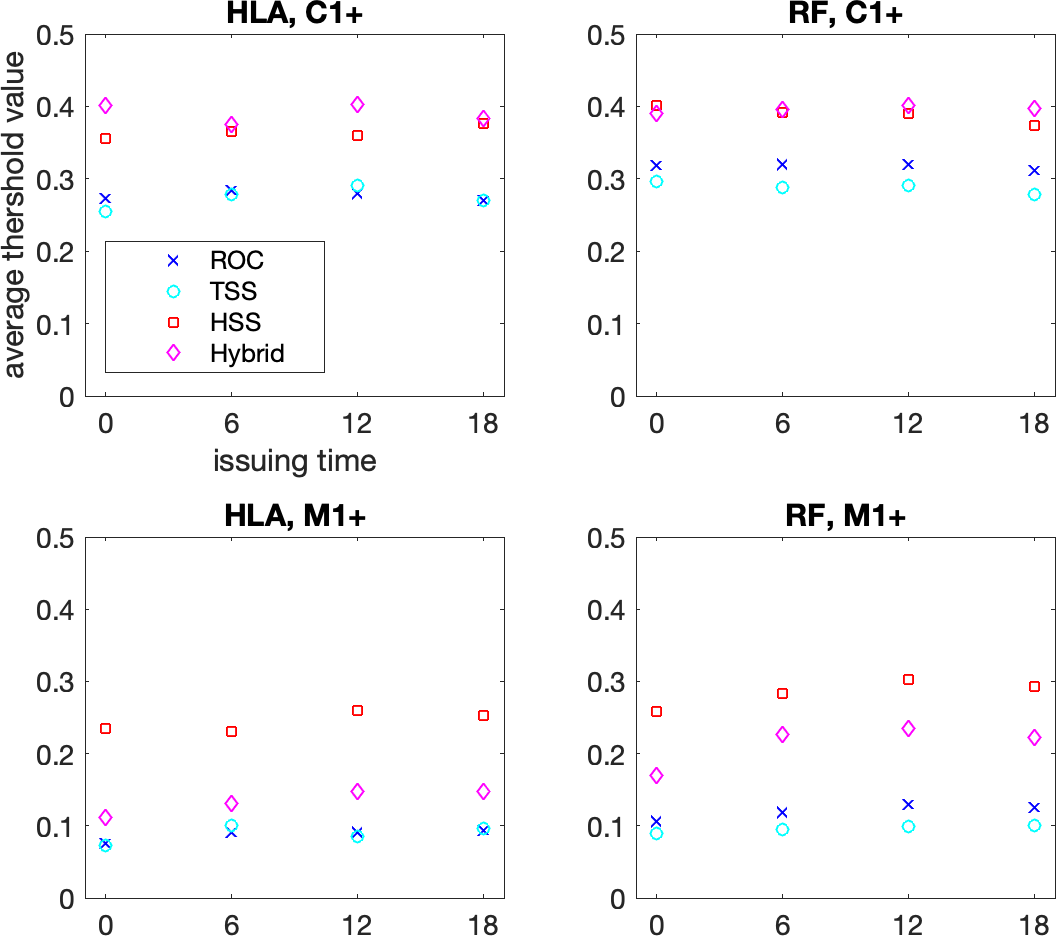}
 \end{center}
 \caption{{\bf{Probability threshold values, $P_{\mathrm{th}}$, averaged over the 100 realizations of the training set. In each panel, symbols indicate the approach applied: hybrid fuzzy clustering (diamonds); HSS optimization (squares); TSS optimization (circles); ROC curve YoudenÕs index optimization (crosses). Top row: prediction of C1+ flares with LASSO and RF (left and right panels, respectively). Bottom row: prediction of M1+flares with LASSO and RF (left and right panels, respectively).}}}
 \label{fig:ROC}
 \end{figure}

The effectiveness of the prediction was assessed by skill scores computed on the previously unseen test sets. Following suggestions in \citet{bletal12}, we chose to use the True Skill Statistic (TSS) and the Heidke Skill Score (HSS), assuming them as representative among skill scores existing in the literature. This said, although not shown here, we performed the analysis using the False Alarm Ratio, Probability of Detection, and Accuracy metrics, obtaining similar results in terms of the relative forecasting effectiveness of the two machine learning methods. {\bf{We point out that all these scores are computed by means of binary predictions applied to the test set. However, as noticed in Section 2.3, LASSO and RF are regression methods providing as outcome real positive numbers that can be interpreted in a probabilistic sense and, in our approach, the transformation of these variables into dichotomous yes/no responses is accomplished by applying a fuzzy clustering technique against the regression outcomes. Other works typically apply an arbitrary probability threshold, $P_{\mathrm{th}}$, of 0.5 to create dichotomous forecasts \citep{leetal19a,leetal19b}, although it should be noted that discriminating thresholds optimized on TSS should find $P_{\mathrm{th}}$ values close to the climatology \citep[i.e., the average flare-day rate;][]{bletal12,baetal16}. Figure \ref{fig:ROC} shows a comparison between the thresholding performances of the clustering technique and the ones provided by the optimization of TSS and HSS and by the use of a ROC curve. Interestingly, the two hybrid regression/clustering approaches provide similar results, which are rather conservative and rather close to the ones achieved by optimizing the HSS, especially in the case of the prediction of C1+ flares. Further, the ROC curve method relies on a cut-off values computed by means of the Youden index \citep{yo50}, which formally leads to the maximization of the TSS; in fact the figure shows that the two values are always very close and the small differences are just related to the different numerical way the thresholding-search schemes were implemented. For C1+ flares, our hybrid approach results in $P_{\mathrm{th}} \approx 0.4$ for both HLA and RF, meaning that our C1+ TSS values are (probably) more comparable to those whose probabilities are converted to dichotomous forecasts using $P_{\mathrm{th}} = 0.5$ (as our $P_{\mathrm{th}}$ lies closer to 0.5 than the average C1+ flare-day rate of $\sim$0.26). The situation is more complex for M1+ flares, however, as the average $P_{\mathrm{th}}$ found by the fuzzy-clustered HLA method is almost equivalent to that optimized on TSS (i.e., approaching the average flare-day rate of 0.04) while for the fuzzy-clustered RF method it instead occupies greater values that lie between the TSS and HSS optimized cases.}}

The averages and standard deviations of the TSS and HSS values over $100$ random realizations of the training/test sets for both prediction methods are shown in Table \ref{table:table_C1_M1}. In the case of HSS, the reliability of average values may be challenged by inconsistent flare/no flare imbalance ratio across the $100$ realizations. However, we have a posteriori checked the sample statistics of the 100 random realizations: average flare/no-flare imbalance ratios across the $100$ test sets are $\sim 0.34$ for C1+ events and $\sim 0.04$ for M1+ events, with relative standard deviations that are $<16\%$ and $<27\%$ of these values, respectively (reflecting the largest relative standard deviations for the four separate UT issuing times considered here).

\begin{figure}
\begin{center}
\begin{tabular}{cc}
 \includegraphics[width=7.5cm]{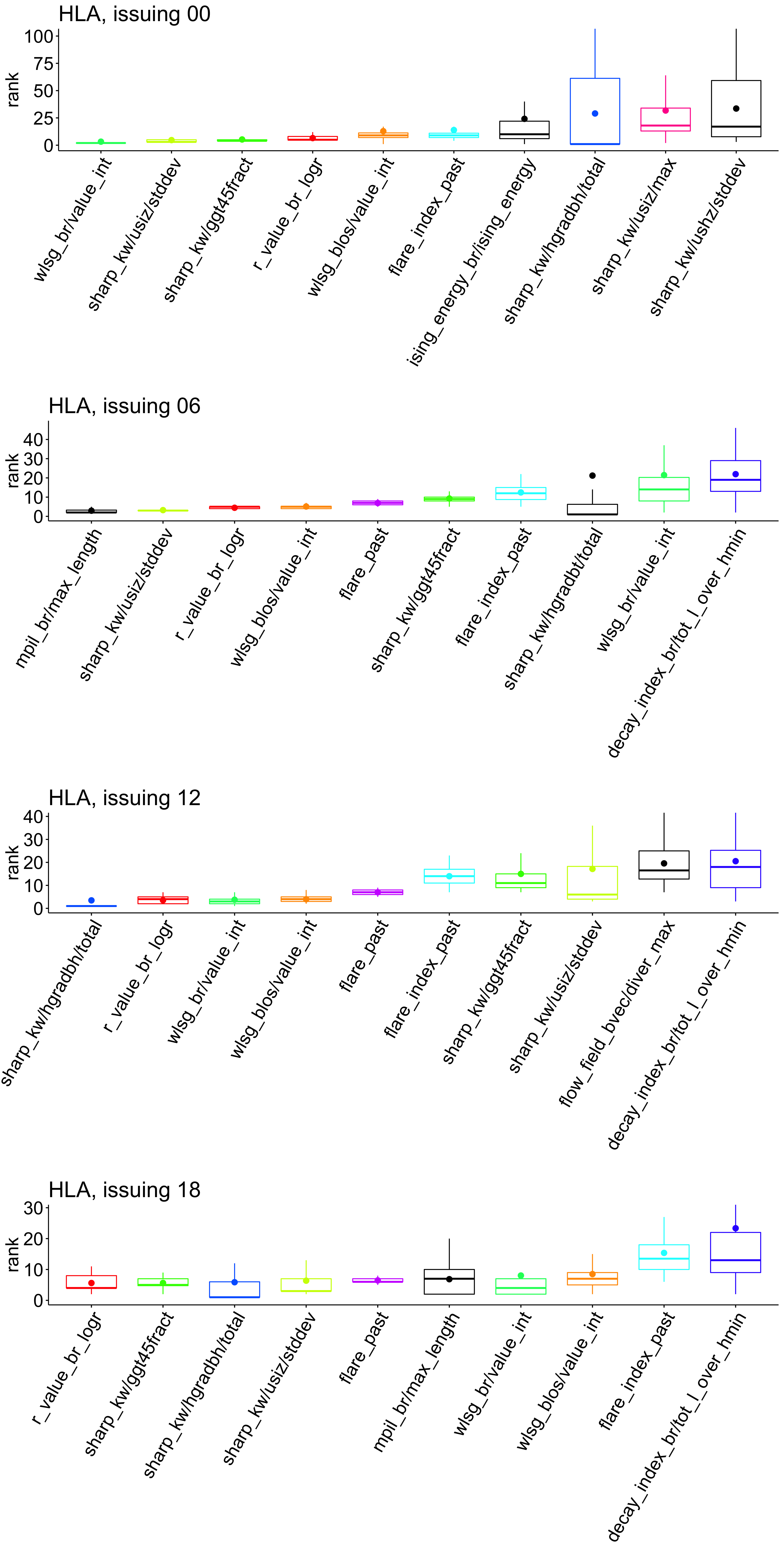} &
 \includegraphics[width=7.5cm]{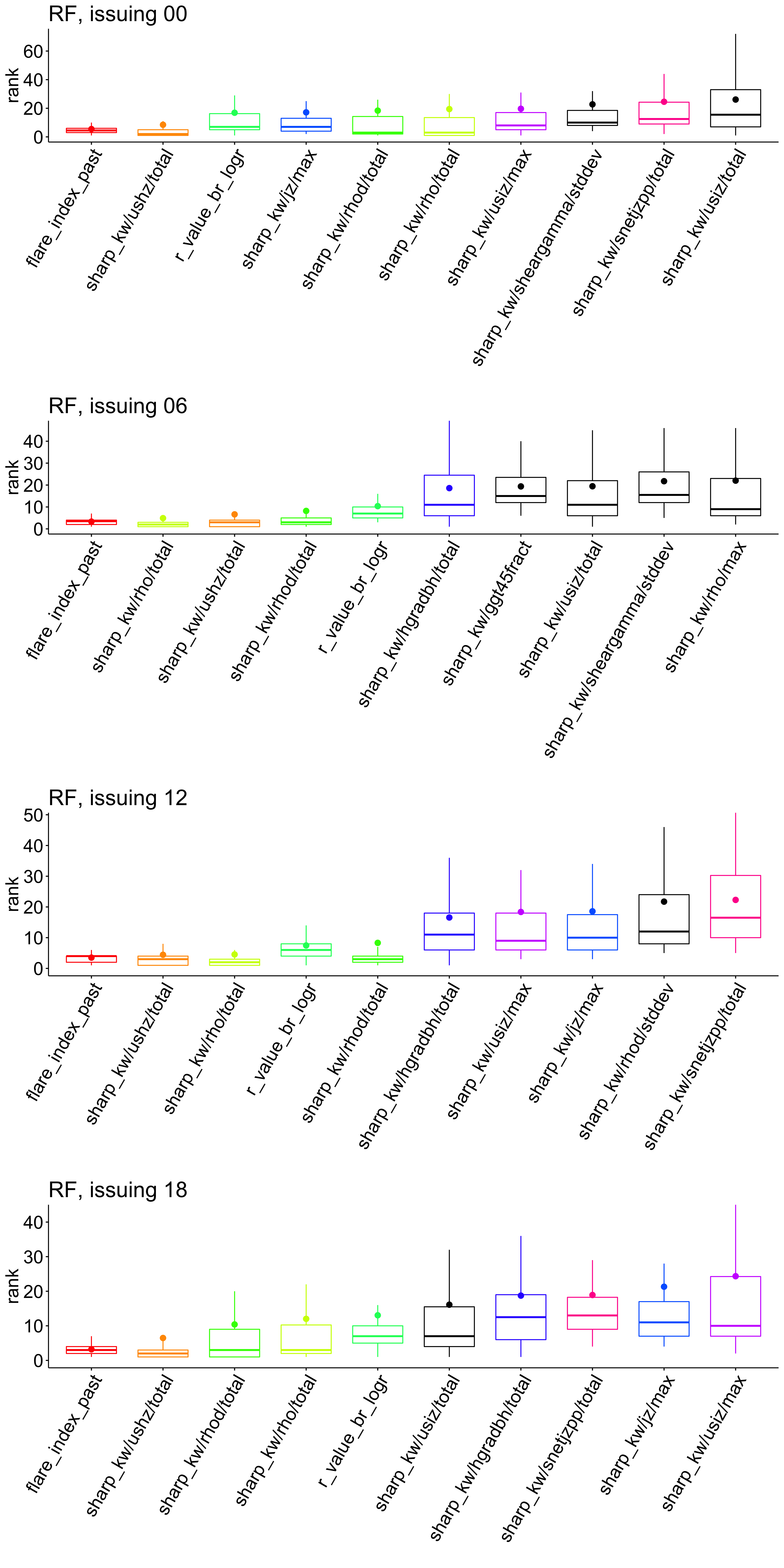}
  \end{tabular}
\end{center}
\caption{Boxplots of the feature ranks provided by RFE as applied against the outcomes of HLA and RF for the $100$ realizations of the training set. The panels show separately the result of the two learning methods (HLA: left column; RF: right column) for the four issuing times considered in the experiment. The focus here is on the prediction of GOES C1+ flares.}
\label{fig:boxplot-1}
 \end{figure}

\begin{figure}
\begin{center}
\begin{tabular}{cc}
 \includegraphics[width=7.5cm]{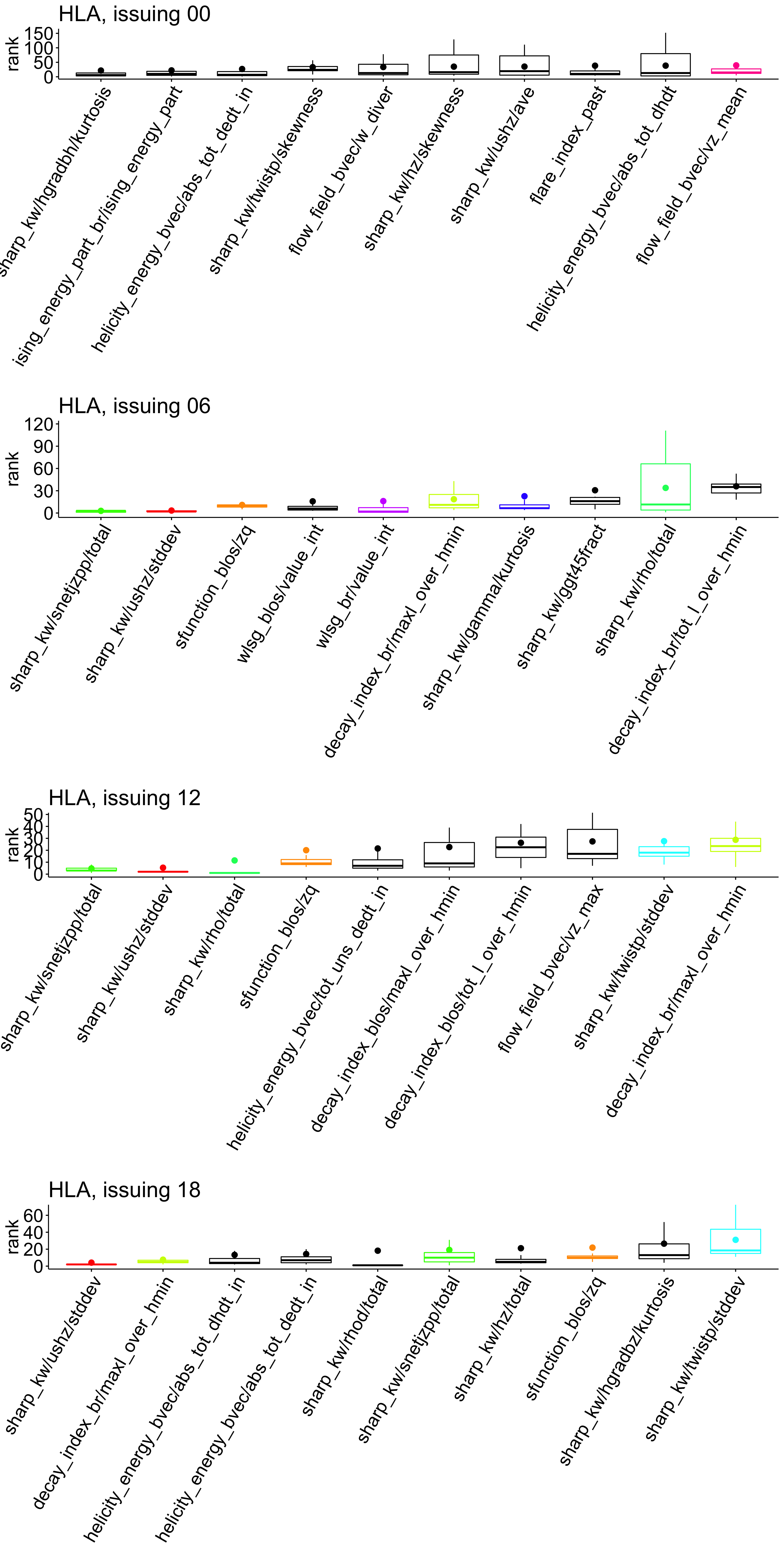} &
 \includegraphics[width=7.5cm]{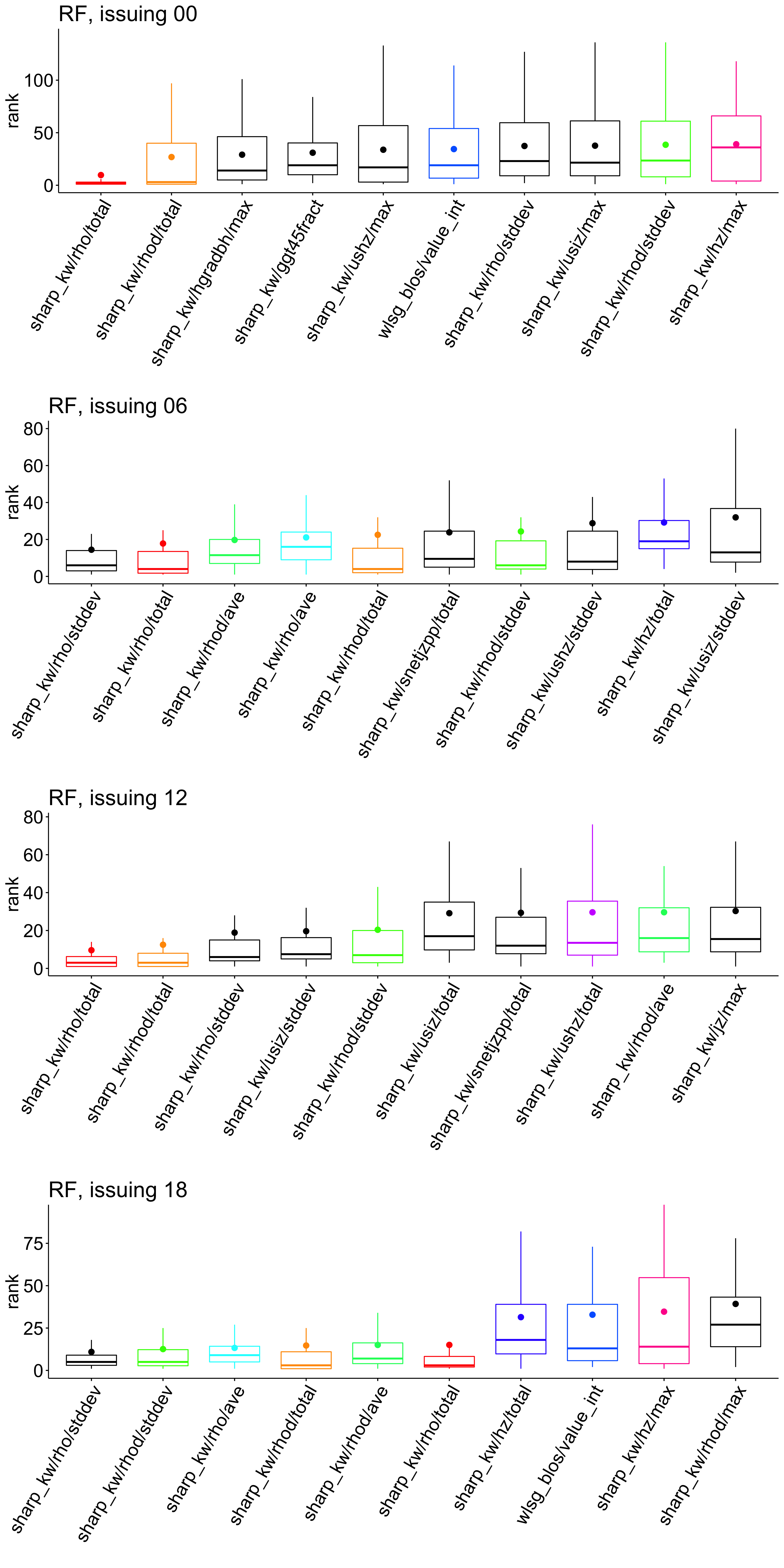}
  \end{tabular}
\end{center}
\caption{ Same as Figure \ref{fig:boxplot-1}, with the focus now being on GOES M1+ flares.}
\label{fig:boxplot-2}
 \end{figure}

Focusing then on the feature ranking process, the boxplots in Figure \ref{fig:boxplot-1} and Figure \ref{fig:boxplot-2} show the top ten features ordered by their mean RFE ranking, obtained by HLA and RF over the $100$ random realizations for each of the four forecast issuing times. Specifically, Figure \ref{fig:boxplot-1} refers to the prediction of GOES C1+ flares while Figure \ref{fig:boxplot-2} refers to the prediction of GOES M1+ flares.

\begin{figure}
\begin{center}
 \includegraphics[width=15.5cm]{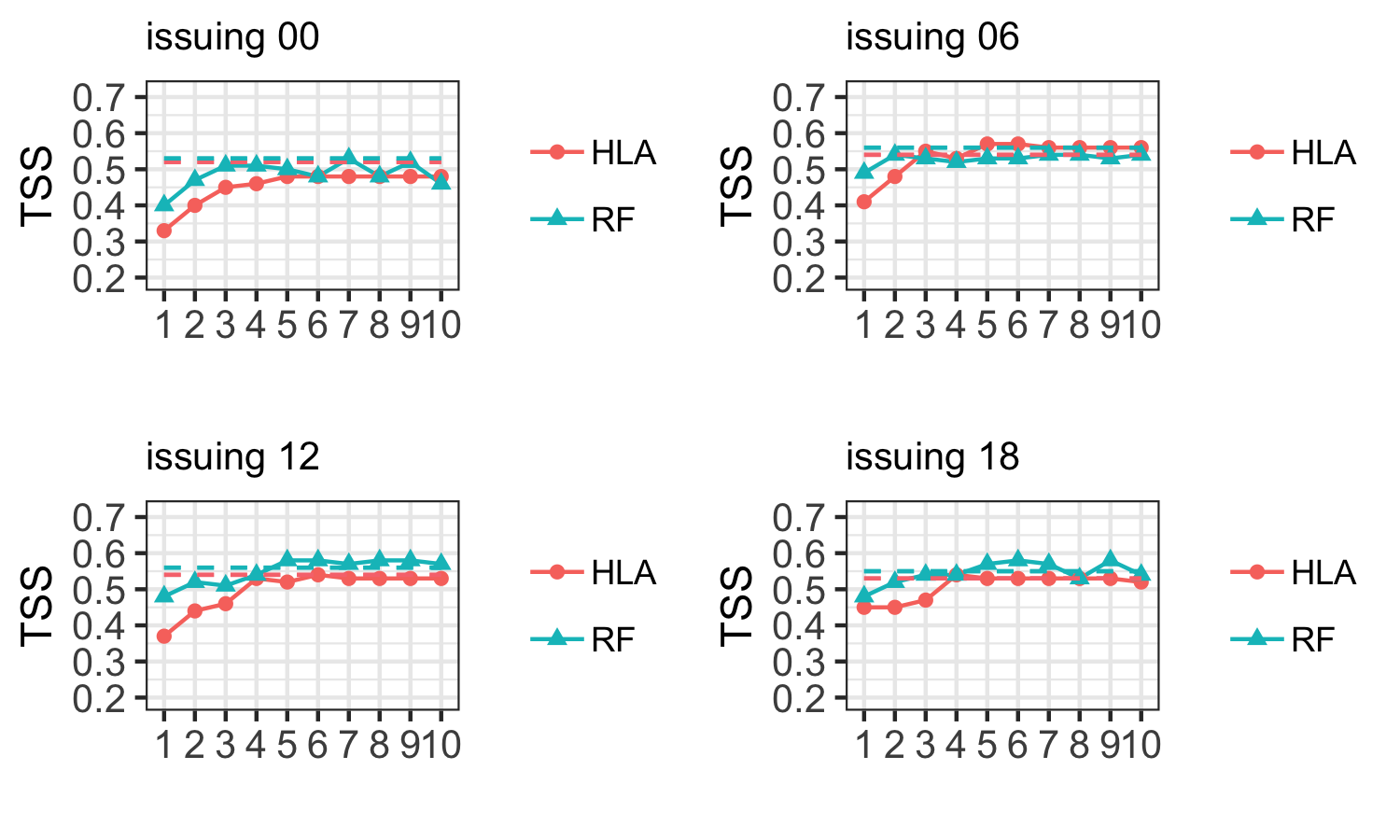}
 \end{center}
\caption{TSS scores obtained by using just the $10$ features with best ranking in decreasing order, from 1 to 10, for both machine learning methods and all four issuing times, in the case of a specific realization of the test set. Features are added one at a time. The plots refer to the prediction of GOES C1+ flares. The dashed horizontal lines are the TSS values obtained by HLA and RF when applied on all $171$ features.}
\label{fig:TSS_C}
 \end{figure}

\begin{figure}
\begin{center}
 \includegraphics[width=15.5cm]{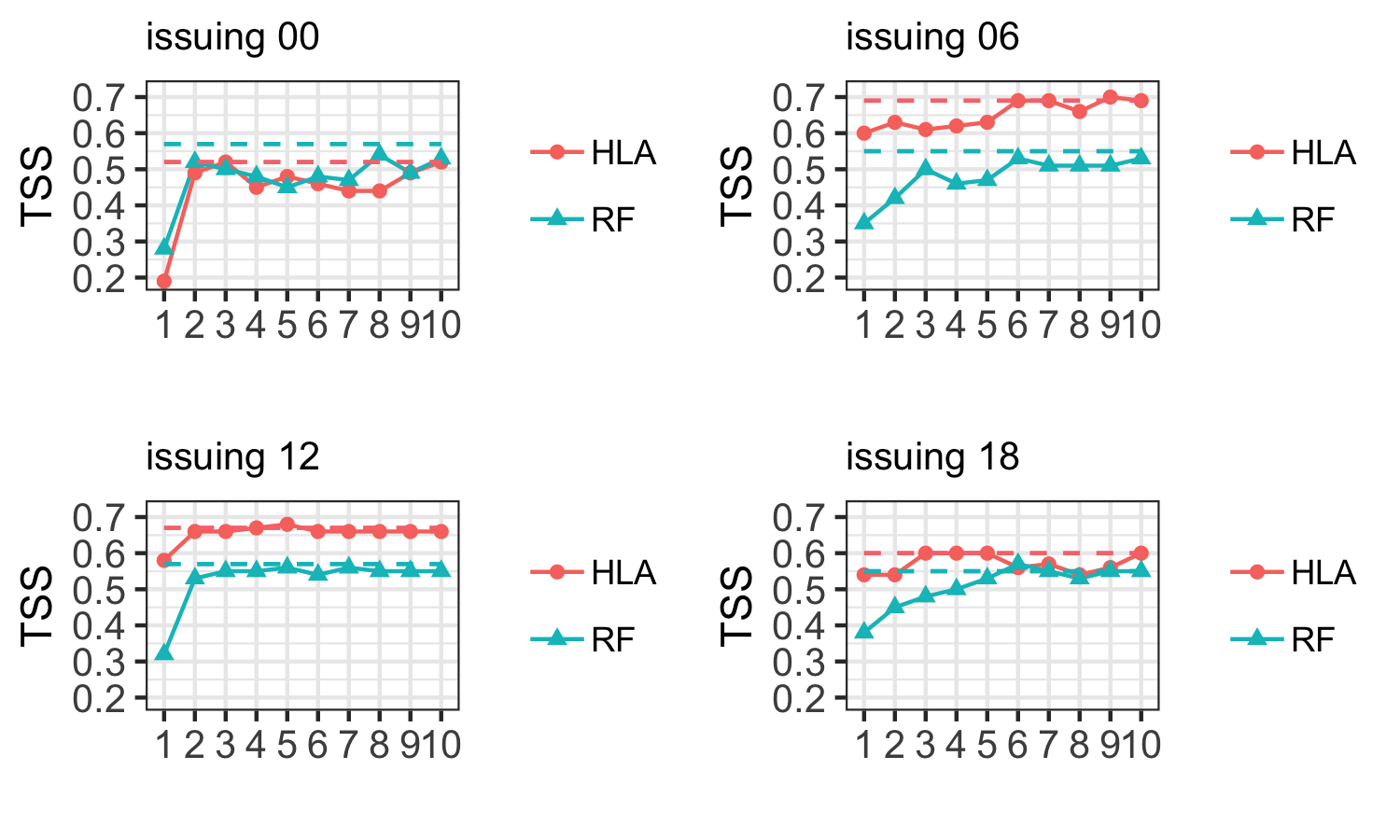}
 \end{center}
\caption{ Same as Figure \ref{fig:TSS_C}, but for GOES M1+ flares.}
\label{fig:TSS_M}
 \end{figure}

From these results it becomes possible to assess the impact of feature selection on the prediction performance, by computing specific skill scores and statistics in a cumulative way. The panels in Figures \ref{fig:TSS_C} and \ref{fig:TSS_M} plot the TSS values obtained by HLA and RF in the case of one specific dataset realization, while adding one feature at a time, starting from the feature with the highest ranking,  down to the feature with the 10th highest ranking. A given feature has the same color throughout each set of plots, for all issuing times.

\begin{figure}
\begin{center}
\begin{tabular}{cc}
 \includegraphics[width=7.5cm]{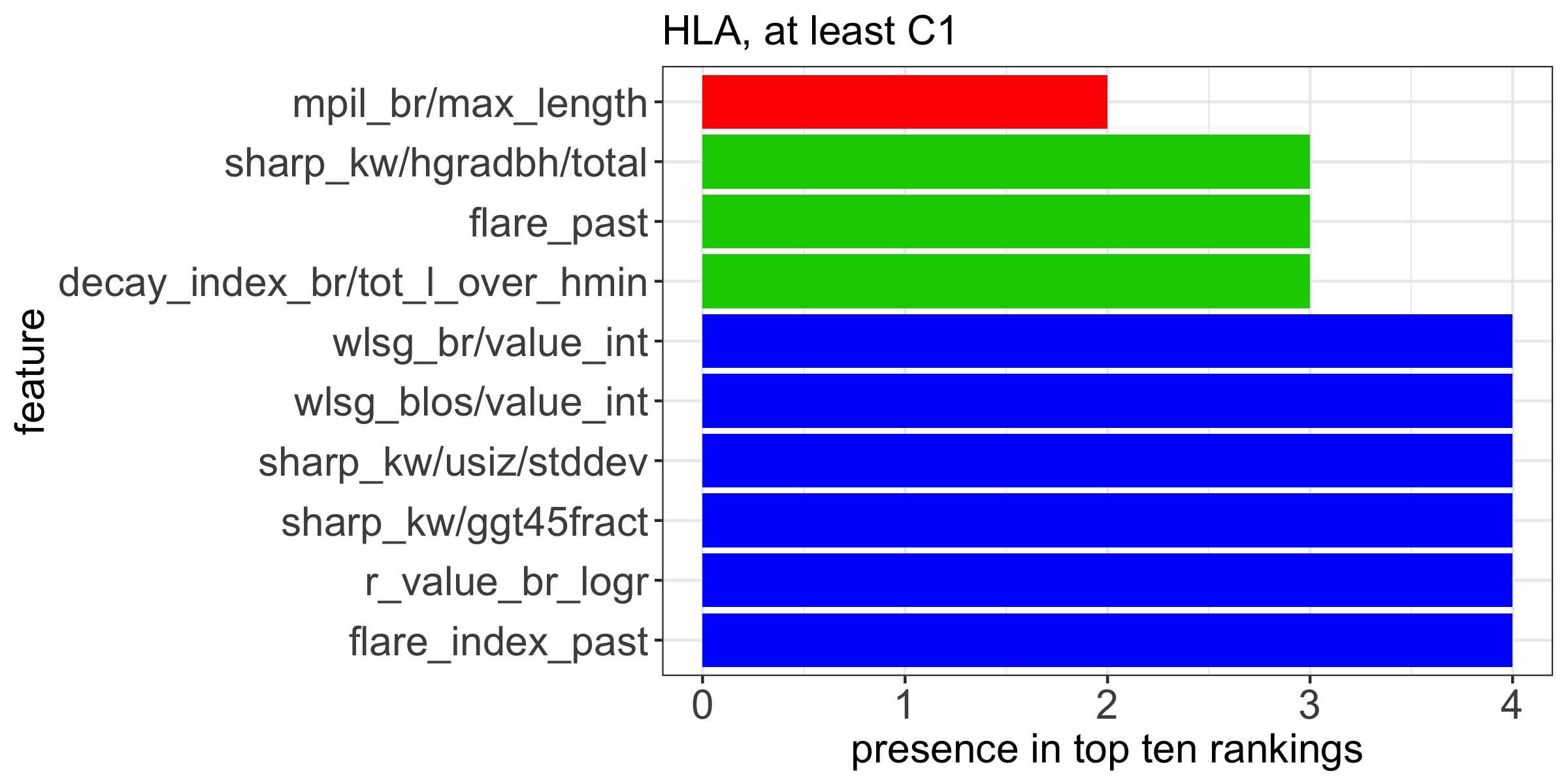} &
 \includegraphics[width=7.5cm]{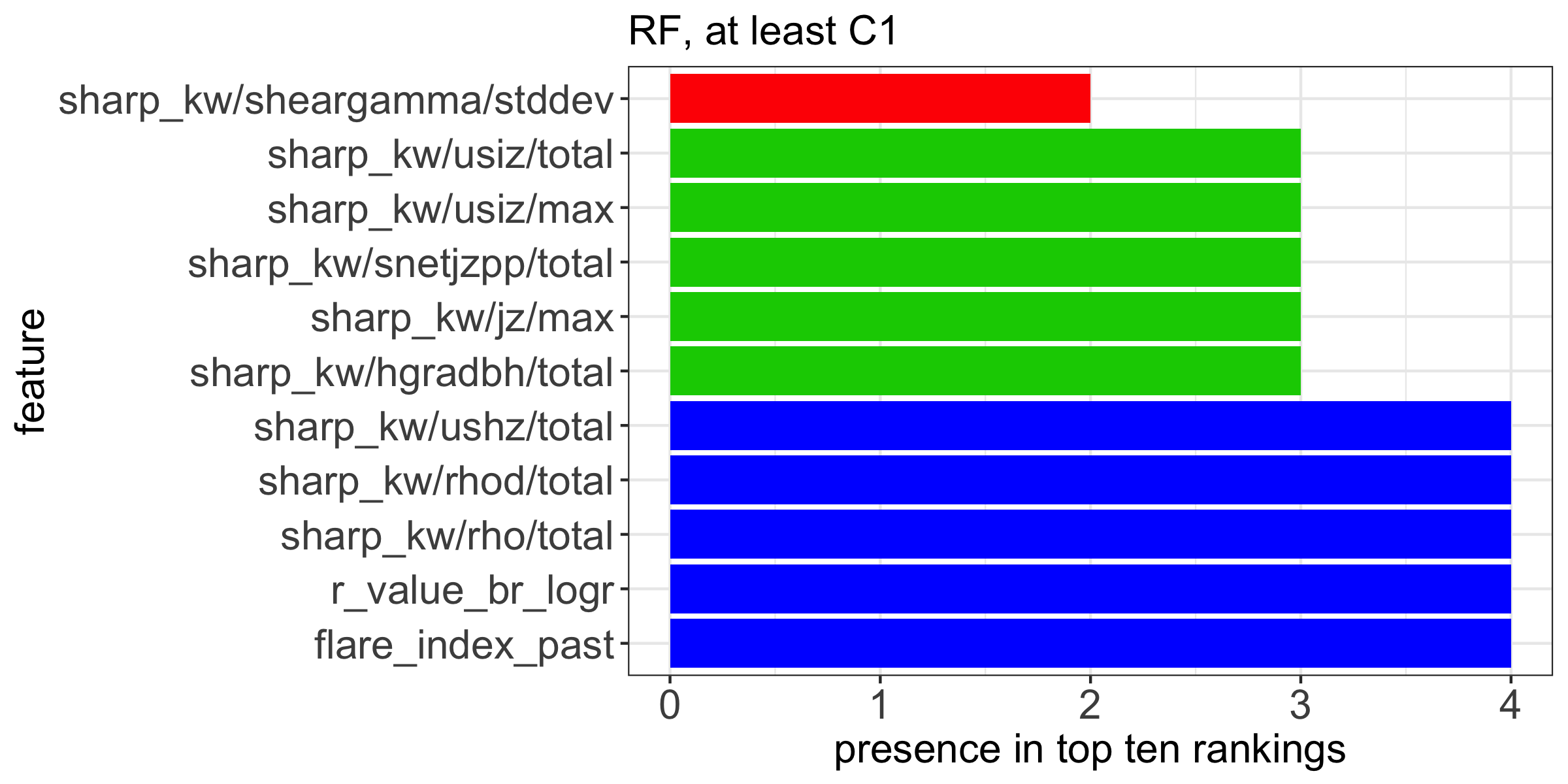} \\
 \includegraphics[width=7.5cm]{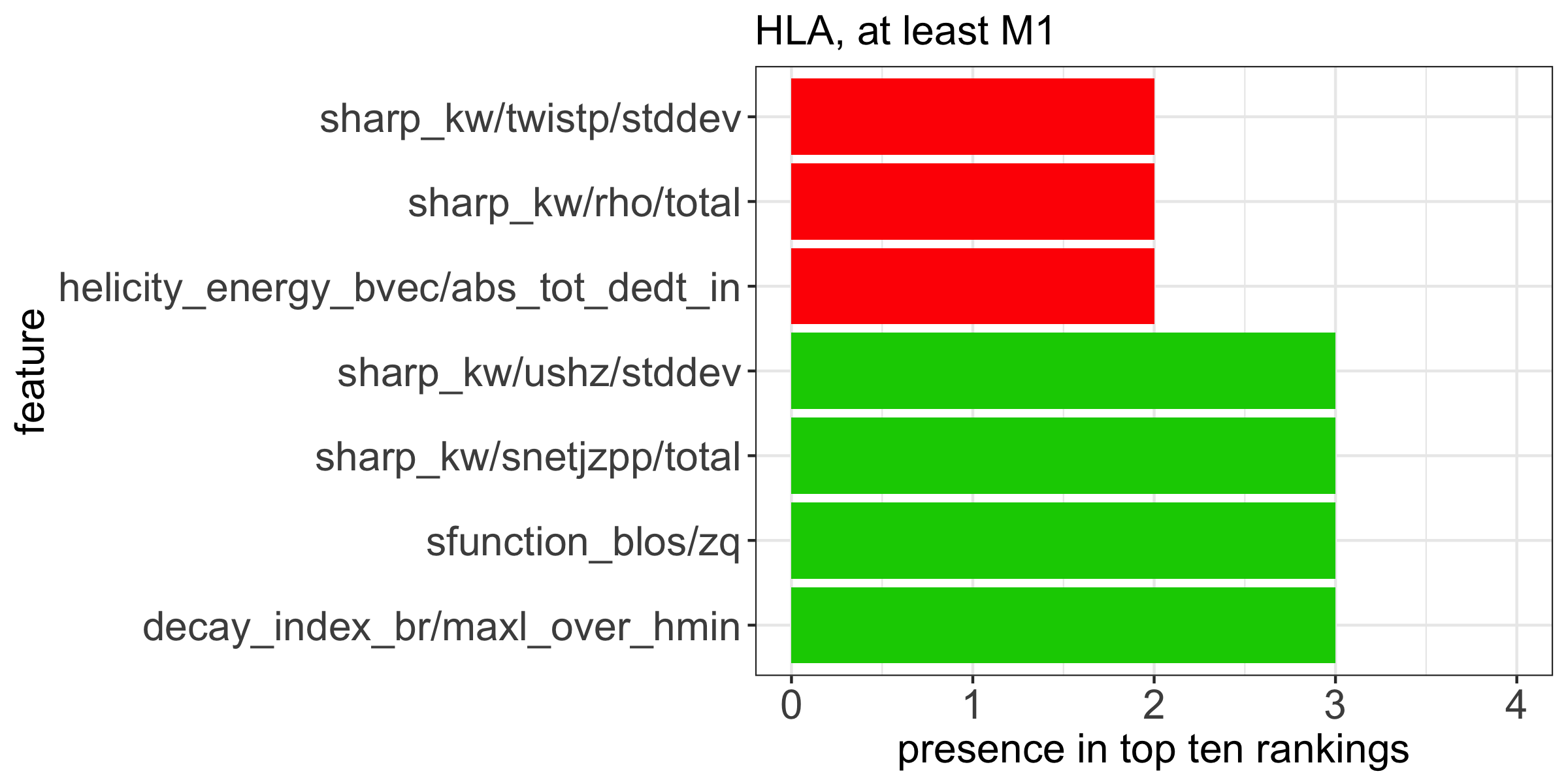} &
 \includegraphics[width=7.5cm]{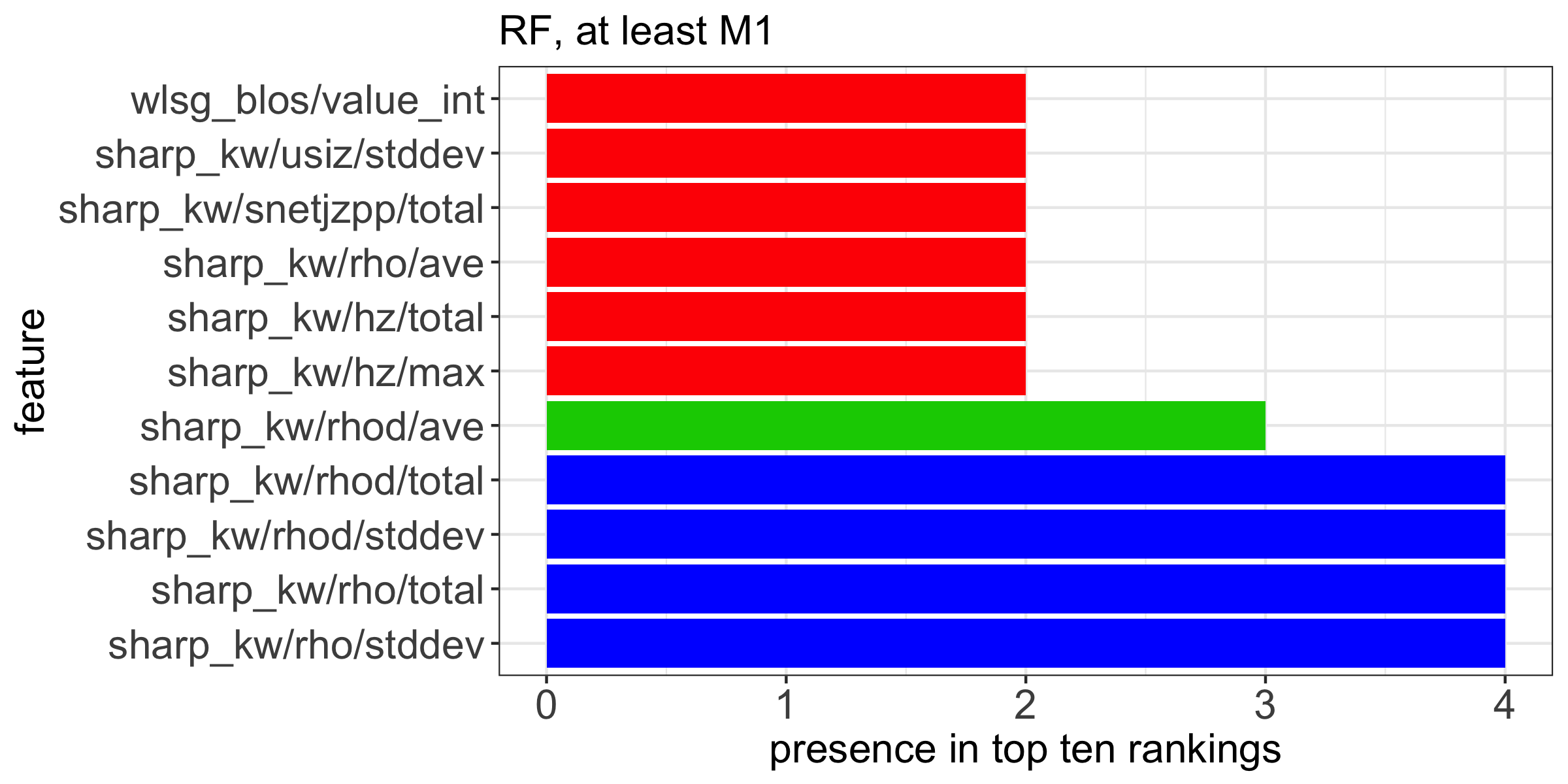} \\
  \end{tabular}
\end{center}
\caption{Histograms counting the number of times each feature is selected in the top-ten rankings, on average over the $100$ random realizations of the test set, for all issuing times and considering both HLA (left column) and RF (right column) as learning machines. Prediction of GOES C1+ flares and GOES M1+ flares is shown in the top and bottom rows, respectively.}
\label{fig:histograms}
 \end{figure}

In order to have a clearer picture of the features that repeatedly show the highest predictive impact, the histograms in Figure \ref{fig:histograms} compute the number of times over the four issuing times that each feature appears in the top-ten ranking of training set averages. These plots only present features that reach the top-ten ranking at least twice out of the four issuing times for a given machine learning method and flare class. For the C1+ flare prediction (Figure \ref{fig:histograms}; top row) one sees, for example, that the past flare history ({\it flare$\_$index$\_$past}) and Schrijver's \citep{sc07} R-value ({\it r$\_$value$\_$br$\_$logr}) consistently  appear for both HLA and RF. This is not the case for the prediction of M1+ flares (Figure \ref{fig:histograms}; bottom row). It should be noted that the importance of the specific features may only be due to the machine learning method used; it is their consistency of appearance, however, that is notable.

\section{Discussion of results}
We first notice that the maximum values of HSS and TSS achieved in Table \ref{table:table_C1_M1} are distinctly different from one, indicating far from perfect performance. Interestingly, these scores are almost systematically smaller than the ones recently achieved by methods illustrated in \citet{boco15} and \citet{fletal18} that use input data with a significantly smaller dimensionality. The methods described in those papers are all supervised, utilize features extracted from {\em{HMI}} data and perform prediction in a 24 h window. However, the way data preparation is performed and, in particular, how the training set is constructed is significantly different than what is done in the present paper. In particular:
\begin{itemize}
\item The test sets utilized in this work to assess the performances of HLA and RF do not contain feature vectors belonging to ARs with feature vectors contained in the training sets. Instead, the training sets utilized in \citet{boco15}, and \citet{fletal18} combined feature vectors belonging to the same ARs in the two sets.
\item We constructed four separate training/testing sets, each corresponding to a specific UT forecast issuing time on all of the days considered. Our results show reasonably consistent forecast performance across these four issuing-time sets. However, the main benefit to this approach is in the interpretation of the feature selection results. Identifying key forecasting features through their appearance in all (or most) of the top-ten feature-ranking lists across these four issuing-time sets increases their robustness through temporal consistency.
\item The training set utilized in \citet{fletal18} is populated with approximately the same number of vectors as the test set, while in our approach (and in the one followed in \citet{boco15}) the machine learning methods are trained with training sets two times more populated than the test sets, which is more realistic with respect to typical experimental settings.
\item Our prediction methods are optimized using a fuzzy clustering technique, while in the other three cases the input parameters are fixed in such a way to optimize a specific skill score (namely, the TSS).
\end{itemize}

In order to assess the impact of these differences against the methods' performance, we trained HLA and RF using all $14931$ point-in-time feature vectors distributing them between training and test sets as was done in \citet{boco15} and \citet{fletal18}. Specifically, we generated the training and test sets focusing on feature vectors instead than on ARs, i.e. we randomly extracted the feature vectors from the database at disposal without imposing any constraint that forbids feature vectors of the same AR to populate both the training and the test set (the two sets are populated as in \citet{boco15}, following a $2:1$ proportion). Further, we did not care for time consistency and so we mixed up feature vectors belonging to different issuing times. Finally, the prediction methods are optimized in such a way to maximize the TSS. {\bf{Table \ref{tab:table-2} shows that TSS increases significantly in the prediction of C1+ and M1+ flares for both HLA and RF and also HSS produced by RF becomes larger, although less significantly. These scores are now more in line with the ones obtained in the other two papers, at the same time showing smaller standard deviations. This leads to the conclusion that, not surprisingly, also in flare prediction the biases introduced in the process of training set generation and the way the algorithms are optimized strongly influence the performance of supervised methods.}}

As far as feature ranking is concerned, it is evident from Figures \ref{fig:boxplot-1} and \ref{fig:boxplot-2} that first, features with the best ranking have the smallest standard deviations, so their impact on prediction is consistently high, regardless of the splitting of the data used for training the machine learning algorithms.  In the case of prediction of GOES C1+ flares, colors largely repeat in all four panels, telling us that features with highest predictive power are common to all issuing times considered. This behavior is not as robust in the case of prediction of GOES M1+ flares, but we are confident that this is a consequence of the lower occurrence rate of M1+ flares and the resulting variation in the flare/no-flare imbalance ratio of the random training sets. A consistent imbalance ratio is more or less guaranteed for C1+ flares, whose comprehensive statistics over solar cycle 24 ensure a well-balanced training process.

Figures \ref{fig:TSS_C} and \ref{fig:TSS_M} show that only a small number of features (up to ten) over the scores of features proposed and/or applied for flare prediction, are sufficient to achieve maximum performance of a given machine learning method. Notice from these figures that the highest-ranking feature alone (feature 1) suffices to give TSS and HSS values that are at least half of the maxima achieved. Up to the 4th feature, the values of TSS and HSS saturate already, indicating that adding more features will not improve (and may, in fact, be detrimental to) prediction performance. Also, provided that flare statistics are sufficient to deal with the flare/no-flare imbalance ratio in the random selection of training and test sets, these few best-performing features are consistent for a given prediction method. {\bf{In their study, \citet{boco15} found that the four most significant features in their analysis were the total unsigned current helicity, total magnitude of the Lorentz force, total photospheric magnetic free energy density, and unsigned vertical current. In our study, Figure \ref{fig:histograms} shows that features associated to the unsigned vertical current (i.e., total, maximum, standard deviation) are among the most temporally consistent of our best-performing features, particularly for C1+ flares (its standard deviation is in the top-ten features of all four UT issuing times for Hybrid LASSO, while its total and maximum are in the top-ten of three of the four UT issuing times for RF) and less so for M1+ flares (its standard deviation is in the top-ten for two of the four UT issuing times). However, the same figure shows that these predictors may well change from method to method, which hampers efforts to understand physically why some features work better than others. Best performers appear to change also for the prediction of different flare classes, which is a very interesting finding that undoubtedly warrants additional investigation in the future.}}  

\section{Conclusions}
This study employs the highest-dimension dataset of prediction features to date in regards to solar flare forecasting, while it shows TSS values similar to the better performing region-by-region forecasting systems in the literature. This point taken, the actual HSS and TSS values are not identical to -- and may even be somewhat lower than -- the respective values reported in \citet{boco15} and in \citet{fletal18}, with the latter study using RF applied to FLARECAST data, as well. The reason is that training and testing of machine learning methods in the present study was not only performed on non-overlapping data as in previous studies, but even the solar active regions selected for training and testing were different. This conclusion seems in line with the considerations contained in \citet{ca19}, {\bf{whose careful assessment of the causality issue identifies this as one of the crucial aspects impacting the forecasting performances.}}

The rationale for using hybrid LASSO and RF in this work is these methods' ability to perform feature ranking via Recursive Feature Elimination, among other methods (i.e., Fisher's score, Gini index, etc.). However, there are 26 machine learning methods implemented in FLARECAST. Their definitive evaluation is in progress, so the values of pertinent skill scores may well increase in future studies utilizing FLARECAST data, in the search for finding the optimal machine learning method(s) for the near-realtime FLARECAST forecasting service. We also understand that a meaningful methodological next step would be to introduce deep learning methods in the pipeline. However, interestingly, the use of these more modern approaches in flare forecasting does not necessarily imply significantly higher skill scores {\bf{(see, e.g. \citep{nietal18}, where TSS and HSS for the prediction of M1+ flares are reported as $0.80$ and $0.26$, respectively)}}.

What will, most likely, not change in the foreseeable future is the following two core conclusions of this work.

First, the current range of properties that have been extracted from the {\em{HMI}} magnetograms show significant redundancy and no more than ten features contained in these properties are sufficient to allow machine learning methods to achieve maximum performance. 

Second, and perhaps foremost, the maximum values of HSS and TSS achieved are distinctly different from one, indicating far from perfect performance. In physical terms, even using the largest flare prediction data volume assembled to date, we have not managed to substantially surpass the performance of a random-chance forecast (as shown by HSS) or to substantially increase the probability of detection ($0.57 - 0.65$ for C1+) despite an encouragingly low probability of false detection ($\sim 0.10$). The latter two compete against each other to result in TSS. This is equivalent to saying that flare prediction remains probabilistic, rather than binary yes/no with a perfect performance. The core reasons for this may be multiple: first, we only rely on photospheric magnetic field data, but flares occur above the line-tied photosphere in the low solar corona. Second, flares may well be intrinsically stochastic phenomena, as adopted in a long-standing working hypothesis \citep{rosva78}, shown conclusively by the flares' time-dependent Poisson waiting times \citep{croetal98, whelit02} and interpretted physically via the concept of self-organized criticality \citep{lh91,lh93,vlaetal95} -- see also \citet{ascetal16} for a comprehensive review. 

%Lifting the stochasticity barrier in predicting solar flare occurrence was a core objective of the FLARECAST project, from its unique vantage point of being able to commit the most comprehensive resources devoted to this purpose to date. Unless a major surprise springs out in the course of evaluation work underway, this barrier will remain in place, as a continuing, key challenge for future groundbreaking machine-learning efforts and data sets (e.g., \citep{angetal19}).

\appendix

%\section{Features extracted from {\em{SDO/AIA}} images}
{\bf{Here we provide details of the FLARECAST feature labels used in this work, with short descriptions and references to their original definition/implementation (or, e.g., detection methods used in their calculation). Features are grouped in the following manner: Table 1 contains those features derived from $B_{\mathrm{los}}$ only and $B_{\mathrm{radial}}$ only \citep{guetal15,heetal08,geru07,zuetal14,ge05,ahetal10,koetal18,maho10,geetal12,sc07,ge12,faetal08}; Table 2 contains those features requiring all three vectormagnetic field components \citep{sc08,kuetal02}; Table 3 contains only those features related to the total and mean quantities provided as the SHARP keywords of \citet{boetal14}.}}

\newpage

\begin{rotatetable}
\begin{deluxetable}{ll}
\tabletypesize{\scriptsize}
\tablecaption{FLARECAST $B\_{\mathrm{los}}$- and $B\_r$-derived feature list with short descriptions.} 
\tablecolumns{3}
\tablewidth{0pt}
%\tablehead{
%\colhead{Feature Label(s)} & \colhead{Description}
%}
\startdata
Feature label & Description \\
\hline\hline 
alpha\_exp\_fft\_blos/alpha, alpha\_exp\_fft\_br/alpha                                     & Fourier power spectral index                                                                          \\
alpha\_exp\_cwt\_blos/alpha, alpha\_exp\_cwt\_br/alpha                                     & Continuous wavelet transform power spectral index                                        \\
\hline
beff\_blos/beff, beff\_br/beff                                                             & Effective connected magnetic field strength $B_{\mathrm{eff}}$                                                    \\
\hline
decay\_index\_blos/max\_l\_over\_hmin               & Max. ratio of MPIL length to min. height of critical decay index  \\
decay\_index\_br/max\_l\_over\_hmin                 & $l/h(n_{\mathrm{cr}})_{\mathrm{min}}$         \\
decay\_index\_blos/tot\_l\_over\_hmin, decay\_index\_br/tot\_l\_over\_hmin                 & Total of all separate MPIL ratios of $l/h(n_{\mathrm{cr}})_{\mathrm{min}}$                  \\
decay\_index\_blos/l\_over\_minhmin, decay\_index\_br/l\_over\_minhmin                     & Ratio of MPIL $l/h(n_{\mathrm{cr}})_{\mathrm{min}}$ (for MPIL having lowest $h(n_{\mathrm{cr}})_{\mathrm{min}}$) \\
decay\_index\_blos/maxl\_over\_hmin, decay\_index\_br/maxl\_over\_hmin                     & Ratio of MPIL $l/h(n_{\mathrm{cr}})_{\mathrm{min}}$ (for longest MPIL)                \\
\hline
flare\_past                                                                                & Binary flag for occurrence of $\geqslant$1 flare in previous 24\,hr                                              \\
flare\_index\_past                                                                         & Accumulated GOES flare peak magnitudes in previous 24\,hr                                                      \\
\hline
frdim\_blos/frdim, frdim\_br/frdim                                                         & Fractal dimension                                                                                                \\
\hline
gs\_slf/g\_s                                                                               & Sum of the horizontal magnetic gradient                                                                          \\
gs\_slf/slf                                                                                & Separation distance lead. and follow. polarity subgroups                                           \\
\hline
ising\_energy\_blos/ising\_energy, ising\_energy\_br/ising\_energy                         & Ising energy (calculated pixel-by-pixel)                                                                        \\
ising\_energy\_part\_blos/ising\_energy\_part, ising\_energy\_part\_br/ising\_energy\_part & Ising energy (calculated using $B_{\mathrm{eff}}$ flux partitions)                                   \\
\hline
lat\_hg                                                                                    & Heliographic latitude of SHARP centroid                                                                        \\
lon\_hg                                                                                    & Heliographic longitude of SHARP centroid                                                                     \\
\hline
mf\_spectrum\_blos/dq, mf\_spectrum\_br/dq                                                 & Multi-fractal generalized correlation dimension spectrum                                                       \\
\hline
mpil\_blos/max\_length, mpil\_br/max\_length                                               & Maximum length of a single MPIL                                                                              \\
mpil\_blos/tot\_length, mpil\_br/tot\_length                                               & Total length of all MPILs                                                                                      \\
mpil\_blos/tot\_usflux, mpil\_br/tot\_usflux                                               & Total unsigned flux around all MPILs                                                                     \\
\hline
r\_value\_blos\_logr, r\_value\_br\_logr                                                   & Schrijver's $R$ ($\log_{10}$ form)                                                                              \\
\hline
sfunction\_blos/zq, sfunction\_br/zq                                                       & Multi-fractal structure function inertial range index                                                          \\
\hline
wlsg\_blos/value\_int, wlsg\_br/value\_int                                                 & Falconer's $^\mathrm{L}\mathrm{WL}_{\mathrm{SG}}$                                                              \\
\hline
\enddata
%\tablenotetext{a}{Reference relates only to the method of MPIL detection.}
\end{deluxetable}
\end{rotatetable}

\newpage
\clearpage

\begin{rotatetable}
\begin{deluxetable}{ll}
\tabletypesize{\scriptsize}
\tablecaption{FLARECAST $B_{r,\theta,\phi}$-derived feature list with short descriptions.}
\tablecolumns{3}
\tablewidth{0pt}
%\tablehead{
%\colhead{Feature Label} & \colhead{Description} & \colhead{Reference}
%}
\startdata
Feature label & Description \\
\hline\hline
flow\_field\_bvec/diver, flow\_field\_bvec/diver\_max, flow\_field\_bvec/diver\_mean          & Flow field divergence (total, maximum, mean)                          \\
\hline
flow\_field\_bvec/shear, flow\_field\_bvec/shear\_max, flow\_field\_bvec/shear\_mean          & Flow field shear (total, maximum, mean)                                    \\
\hline
flow\_field\_bvec/v\_ mean, flow\_field\_bvec/v\_median                                       & Flow field total velocity magnitude (mean, median)                         \\
\hline
flow\_field\_bvec/vz\_max, flow\_field\_bvec/vz\_mean                                         & Flow field vertical velocity magnitude (mean, median)                       \\
\hline
flow\_field\_bvec/w\_diver, flow\_field\_bvec/w\_diver\_max, flow\_field\_bvec/w\_diver\_mean & Flux-weighted flow field divergence (total, maximum, mean)         \\
\hline
flow\_field\_bvec/w\_shear, flow\_field\_bvec/w\_shear\_max, flow\_field\_bvec/w\_shear\_mean & Flux-weighted flow field shear (total, maximum, mean)              \\
\hline
helicity\_energy\_bvec/abs\_tot\_dedt                                                         & Abs. val. net vertical Poynting flux                              \\
helicity\_energy\_bvec/abs\_tot\_dedt\_in                                                     & Abs. val. net vertical Poynting flux (emerg. comp.)         \\
helicity\_energy\_bvec/abs\_tot\_dedt\_sh                                                     & Abs. val. net vertical Poynting flux (shear. comp.)          \\
helicity\_energy\_bvec/abs\_tot\_dedt\_in\_plus\_sh                                           & Emerg. $+$ shear. abs. values net vertical Poynting flux  \\
\hline
helicity\_energy\_bvec/abs\_tot\_dhdt                                                         & Abs. val. net vertical helicity flux                               \\
helicity\_energy\_bvec/abs\_tot\_dhdt\_in                                                     & Abs. val. net vertical helicity flux (emerg. comp.)          \\
helicity\_energy\_bvec/abs\_tot\_dhdt\_sh                                                     & Abs. val. net vertical helicity flux (emerg. comp.)          \\
helicity\_energy\_bvec/abs\_tot\_dhdt\_in\_plus\_sh                                           & Emerg. $+$ shear. abs. values net vertical helicity flux \\
\hline
helicity\_energy\_bvec/tot\_uns\_dedt                                                         & Total unsigned vertical Poynting flux                                       \\
helicity\_energy\_bvec/tot\_uns\_dedt\_in                                                     & Tot. unsign. vertical Poynting flux (emerg. comp.)                \\
helicity\_energy\_bvec/tot\_uns\_dedt\_sh                                                     & Tot. unsign. vertical Poynting flux (shear. comp.)                 \\
\hline
helicity\_energy\_bvec/tot\_uns\_dhdt                                                         & Tot. unsign. vertical helicity flux                                      \\
helicity\_energy\_bvec/tot\_uns\_dhdt\_in                                                     & Tot. unsign. vertical helicity flux (emerg. comp.)                \\
helicity\_energy\_bvec/tot\_uns\_dhdt\_sh                                                     & Tot. unsign. vertical helicity flux (shear. comp.)                  \\
\hline
nn\_currents/tot\_us\_cur                                                                  & Total unsigned non-neutralized currents                                                                         \\
\hline
\enddata
\end{deluxetable}
\end{rotatetable}

\newpage

\begin{rotatetable}
\begin{deluxetable}{ll}
\tabletypesize{\tiny}
\tablecaption{FLARECAST SHARP-keyword related feature list with short descriptions.} 
\tablecolumns{2}
\tablewidth{0pt}
%\tablehead{
%\colhead{Feature Label(s)} & \colhead{Description}
%}
\startdata
Feature label & Description \\
\hline\hline
sharp\_kw/gamma/ave, sharp\_kw/gamma/stddev                     & Field inclin. ang.  (mean, st. dev.)\\
sharp\_kw/gamma/skewness, sharp\_kw/gamma/kurtosis                     & Field inclin. ang.  (skewn., kurt.)\\
sharp\_kw/gamma/total, sharp\_kw/gamma/max, sharp\_kw/gamma/median                                         & Field inclin. ang. (tot., max., med.)\\
\hline
sharp\_kw/ggt45fract                                                                                                & $\%$ tot. area with shear angle $>$ 45$^{\circ}$\\
\hline
sharp\_kw/hgradbh/ave, sharp\_kw/hgradbh/stddev             & Horiz. grad. $B_{\mathrm{hor}}$ (mean, st. dev.)\\
sharp\_kw/hgradbh/skewness, sharp\_kw/hgradbh/kurtosis             & Horiz. grad. $B_{\mathrm{hor}}$ (skewn., kurt.)\\
sharp\_kw/hgradbh/total, sharp\_kw/hgradbh/max, sharp\_kw/hgradbh/median                                            & Horiz. grad. $B_{\mathrm{hor}}$ (tot, max, med)\\
\hline
sharp\_kw/hgradbt/ave, sharp\_kw/hgradbt/stddev             & Horiz. grad. $B_{\mathrm{tot}}$ (mean, st. dev.)\\
sharp\_kw/hgradbt/skewness, sharp\_kw/hgradbt/kurtosis             & Horiz. grad. $B_{\mathrm{tot}}$ (skewn., kurt.)\\
sharp\_kw/hgradbt/total, sharp\_kw/hgradbt/max, sharp\_kw/hgradbt/median                                            & Horiz. grad. $B_{\mathrm{tot}}$ (tot., max., med.)\\
\hline
sharp\_kw/hgradbz/ave, sharp\_kw/hgradbz/stddev             & Horiz. grad. $B_{r}$ (mean, st. dev.)\\
sharp\_kw/hgradbz/skewness, sharp\_kw/hgradbz/kurtosis             & Horiz. grad. $B_{r}$ (skewn., kurt.)\\
sharp\_kw/hgradbz/total, sharp\_kw/hgradbz/max, sharp\_kw/hgradbz/median                                            & Horiz. grad. $B_{r}$ (tot., max., med.)\\
\hline
sharp\_kw/hz/ave, sharp\_kw/hz/stddev                               & Vert. curr. hel. (mean, st. dev.)\\
sharp\_kw/hz/skewness, sharp\_kw/hz/kurtosis                                 & Vert. curr. hel. (skewn., kurt.)\\
sharp\_kw/hz/total, sharp\_kw/hz/max, sharp\_kw/hz/median                                                           & Vert. curr. hel. (tot., max., med.)\\
\hline
sharp\_kw/jz/ave, sharp\_kw/jz/stddev                          & Vert. curr. (mean, st. dev.)\\
sharp\_kw/jz/skewness, sharp\_kw/jz/kurtosis                                 & Vert. curr. (skewn., kurt.)\\
sharp\_kw/jz/total, sharp\_kw/jz/max, sharp\_kw/jz/median                                                           & Vert. curr. (tot., max., med.)\\
\hline
sharp\_kw/rho/ave, sharp\_kw/rho/stddev                             & Photosph. excess magn. en. (mean, st. dev.)\\
sharp\_kw/rho/skewness, sharp\_kw/rho/kurtosis                             & Photosph. excess magn. en. (skewn., kurt.)\\
sharp\_kw/rho/total, sharp\_kw/rho/max, sharp\_kw/rho/median                                                        & Photosph. excess magn. en. (tot., max., med.)\\
\hline
sharp\_kw/rhod/ave, sharp\_kw/rhod/stddev                        & Photosph. excess magn. en. dens. (mean, st. dev.)\\
sharp\_kw/rhod/skewness, sharp\_kw/rhod/kurtosis                         & Photosph. excess magn. en. dens. (skewn., kurt.)\\
sharp\_kw/rhod/total, sharp\_kw/rhod/max, sharp\_kw/rhod/median                                                     & Photosph. excess magn. en. dens (tot., max., med.)\\
\hline
sharp\_kw/sflux/ave, sharp\_kw/sflux/stddev                    & Signed flux ( mean, st. dev.)\\
sharp\_kw/sflux/skewness, sharp\_kw/sflux/kurtosis                     & Signed flux (skewn., kurt.)\\
sharp\_kw/sflux/total, sharp\_kw/sflux/max, sharp\_kw/sflux/median                                                  & Signed flux (tot., max., med.)\\
\hline
sharp\_kw/sheargamma/ave, sharp\_kw/sheargamma/stddev  & $B_{\mathrm{tot}}$ shear angle (mean, st. dev.)\\
sharp\_kw/sheargamma/skewness, sharp\_kw/sheargamma/kurtosis & $B_{\mathrm{tot}}$ shear angle (skewn., kurt.)\\
sharp\_kw/sheargamma/total, sharp\_kw/sheargamma/max, sharp\_kw/sheargamma/median                                   & $B_{\mathrm{tot}}$ shear angle (tot., max., med.)\\
\hline
sharp\_kw/snetjzpp/total                                                                                            & Sum abs. val. net currents per polarity\\
\hline
sharp\_kw/twistp/ave, sharp\_kw/twistp/stddev, sharp\_kw/twistp/skewness, sharp\_kw/twistp/kurtosis                 & Twist parameter (mean, st. dev., skewn., kurt.)\\
sharp\_kw/twistp/total, sharp\_kw/twistp/max, sharp\_kw/twistp/median                                               & Twist parameter (tot., max., med.)\\
\hline
sharp\_kw/usflux/ave, sharp\_kw/usflux/stddev, sharp\_kw/usflux/skewness, sharp\_kw/usflux/kurtosis                 & Uns. flux (mean, st. dev., skewn., kurt.)\\
sharp\_kw/usflux/total, sharp\_kw/usflux/max, sharp\_kw/usflux/median                                               & Uns. flux (tot., max., med.)\\
\hline
sharp\_kw/ushz/ave, sharp\_kw/ushz/stddev, sharp\_kw/ushz/skewness, sharp\_kw/ushz/kurtosis                         & Uns. vert. curr. hel. (mean, st. dev., skewn., kurt.)\\
sharp\_kw/ushz/total, sharp\_kw/ushz/max, sharp\_kw/ushz/median                                                     & Uns. vert. curr. hel. (tot., max., med.)\\
\hline
sharp\_kw/usiz/ave, sharp\_kw/usiz/stddev, sharp\_kw/usiz/skewness, sharp\_kw/usiz/kurtosis                         & Uns. vert. curr. (mean, st. dev., skewn., kurt.)\\
sharp\_kw/usiz/total, sharp\_kw/usiz/max, sharp\_kw/usiz/median                                                     & Uns. vert. curr. (tot., max., med.)\\
\hline
\enddata
\end{deluxetable}
\end{rotatetable}

\newpage

\noindent {\bf{Acknowledgements.}}
This study was enabled by the European Union's Horizon 2020 Research and Innovation Action Flare Likelihood And Region Eruption foreCASTing (FLARECAST) project, under grant agreement No. 640216.
All authors warmly acknowledge the support of the FLARECAST project.\\

\newpage

\begin{table}
\centering
\caption{ Average TSS- and HSS-values, along with applicable standard deviations, over the outcomes of HLA and RF as applied against $100$ random realizations of the training/test sets.}\label{table:table_C1_M1}
%\medskip
%\rotatebox{90}
\begin{tabular}{|c||c|c||c|c|}
%\hline
% & \multicolumn{2}{c||}{Training set  - {\bf C1+} flares} & \multicolumn{2}{c|}{Training set  - {\bf M1+} flares}\\
%\hline 
% 00:00:00UT & TSS & HSS & TSS & HSS \\
%\hline
%HLA &	0.51 $\pm$ 0.02&	0.54 $\pm$ 0.02&	  0.64 $\pm$ 0.05	& 0.31 $\pm$ 0.06  \\
%RF	&	0.66 $\pm$ 0.03&	0.71 $\pm$ 0.02&	 0.19 $\pm$ 0.10	& 0.30  $\pm$ 0.14   \\
%\hline
% 06:00:00UT & TSS & HSS &  TSS & HSS \\
%\hline
%HLA &	0.55 $\pm$ 0.01&	0.55 $\pm$ 0.01&	0.69 $\pm$ 0.03	& 0.37 $\pm$ 0.04  \\
%RF	&	0.67 $\pm$ 0.02&	0.71 $\pm$ 0.01&	0.37 $\pm$ 0.10	& 0.51  $\pm$ 0.06  \\
%\hline
%12:00:00UT  & TSS & HSS & TSS & HSS \\
%\hline
%HLA &	0.52 $\pm$ 0.02&	0.55 $\pm$ 0.01&	0.68 $\pm$ 0.02	& 0.40 $\pm$ 0.05  \\
%RF	&	0.67 $\pm$ 0.02&	0.71 $\pm$ 0.01&	0.40 $\pm$ 0.05	& 0.64 $\pm$ 0.05   \\
%\hline
% 18:00:00UT& TSS & HSS & TSS & HSS \\
%\hline
%HLA &	0.54 $\pm$ 0.02&	0.56 $\pm$ 0.02&	 0.68 $\pm$ 0.03	& 0.41 $\pm$ 0.04  \\
%RF	&	0.66 $\pm$ 0.02&	0.71 $\pm$ 0.02&	0.38 $\pm$ 0.06	& 0.51  $\pm$ 0.06   \\
%\hline
\hline
 & Test set  - C1+ & Test set  - C1+ & Test set - M1+ & Test set - M1+ \\
\hline 
00:00:00UT & TSS & HSS &  TSS & HSS \\
\hline
HLA & 0.48 $\pm$ 0.06&	0.51 $\pm$ 0.05&	 0.56 $\pm$ 0.14	& 0.27 $\pm$ 0.06  \\
RF  &	0.53 $\pm$ 0.05&	0.52 $\pm$ 0.04&	 0.48 $\pm$ 0.14	& 0.33 $\pm$ 0.09	\\
\hline
06:00:00UT & TSS & HSS & TSS & HSS \\
\hline
HLA & 0.53 $\pm$ 0.03&	0.54 $\pm$ 0.03&	 0.67 $\pm$ 0.05	& 0.35 $\pm$ 0.04 \\
RF  &0.54 $\pm$ 0.03&	0.54 $\pm$ 0.03&	 0.49 $\pm$ 0.08	& 0.42 $\pm$ 0.06	\\
\hline
12:00:00UT  & TSS & HSS & TSS & HSS \\
\hline
HLA &	0.51 $\pm$ 0.04&	0.54 $\pm$ 0.03&	 0.66 $\pm$ 0.06	& 0.38 $\pm$ 0.04  \\
RF  &	0.53 $\pm$ 0.03&	0.53 $\pm$ 0.03&	 0.51 $\pm$ 0.09	& 0.43 $\pm$ 0.06	\\
\hline
18:00:00UT & TSS & HSS &  TSS & HSS \\
\hline
HLA & 0.54 $\pm$ 0.04&	0.55 $\pm$ 0.03&	  0.64 $\pm$ 0.07	& 0.39 $\pm$ 0.04  \\
RF  &	0.55 $\pm$ 0.03&	0.55$\pm$ 0.03&	 0.53 $\pm$ 0.09	& 0.43 $\pm$ 0.06	\\
\hline
\end{tabular}
\end{table}

\begin{table}
\centering
\caption{Average TSS- and HSS-values, along with applicable standard deviations, over the outcomes of HLA and RF as applied against $100$ random realizations of the training/test sets. The training sets have been generated according to the same procedure as in \citet{boco15,fletal18}. The scores presented in those papers are reported in this Table.}\label{tab:table-2}
%\medskip
%\rotatebox{90}
\begin{tabular}{|c||c|c||c|c|}
\hline
 & Test set  - C1+ & Test set  C1+ & Test set - M1+ & Test set - M1+\\
\hline 
& TSS & HSS & TSS & HSS \\
\hline
HLA & $0.58 \pm 0.01$ & $0.51 \pm 0.01$ & $0.70 \pm 0.02$ & $0.31 \pm 0.03$ \\
RF & $0.61 \pm 0.01$ & $0.56 \pm 0.02$ & $0.71 \pm 0.03$ & $0.39 \pm 0.02$ \\
Florios et al (2018) & $0.60 \pm 0.01$ & $0.59 \pm 0.01$ & $0.74 \pm 0.02$ & $0.49 \pm 0.01$ \\
Bobra \& Couvidat (2015) & $\ldots$ & $\ldots$ & $0.76 \pm 0.04$ & $0.52 \pm 0.04$ \\
\hline
\end{tabular}
\end{table}

\end{document}